\documentclass[prl,twocolumn,superscriptaddress,groupedaddress,tightenlines,eqsecnum]{revtex4-1}
\usepackage[T1]{fontenc}
\usepackage{mathtools}
\usepackage{amssymb}
\usepackage{stmaryrd}
\usepackage[mathscr]{euscript}
\usepackage{dsfont}
\usepackage{slashed}
\usepackage{currvita}
\usepackage[integrals]{wasysym}

\DeclareRobustCommand{\soint}{\mathbin{\!\int\hspace{-8.5pt}\text{\large{$\circ$}}}}

\usepackage[colorlinks=true,linkcolor=blue,citecolor=magenta,linktocpage=true]{hyperref}
\usepackage{braket}
\usepackage{titlesec}
\titleformat*{\section}{\normalsize\bfseries}
\titleformat*{\subsection}{\normalsize\bfseries}
\titleformat*{\subsubsection}{\normalsize\bfseries}
\makeatletter
\renewcommand{\@dotsep}{1000}

\def\be#1\ee{\begin{align}#1\end{align}}
\def\bsub#1\esub{\begin{subequations}#1\end{subequations}}

\def\q{\qquad}
\def\f{\frac}
\def\eps{\varepsilon}

\def\lb{\big\lbrace}
\def\rb{\big\rbrace}

\def\ip{\raisebox{0.03cm}{\text{$\lrcorner\,$}}}

\def\para{\text{\tiny{$\parallel$}}}

\def\de_\omega{\mathrm{D}}
\def\de{\mathrm{d}}
\def\i{\mathrm{i}}

\def\A{\mathcal{A}}

\def\D{\mathcal{D}}
\def\E{\mathcal{E}}
\def\F{\mathcal{F}}
\def\G{\mathcal{G}}

\def\J{\mathcal{J}}

\def\L{\mathcal{L}}

\def\O{\mathcal{O}}
\def\P{\mathcal{P}}
\def\Q{\mathcal{Q}}

\def\T{\mathcal{T}}

\def\pe{\phantom{\ =}}
\newcommand{\ds}{\displaystyle}

\begin{document}

\title{\large{\textbf{\sffamily Dual diffeomorphisms and finite distance asymptotic symmetries in 3d gravity}}}
\author{\normalsize{\sffamily\vspace{-0.2cm}Marc Geiller \& Christophe Goeller}}
\affiliation{\vspace{-0.3cm}Univ Lyon, ENS de Lyon, Univ Claude Bernard Lyon 1,\\ CNRS, Laboratoire de Physique, UMR 5672, F-69342 Lyon, France}

\begin{abstract}
We study the finite distance boundary symmetry current algebra of the most general first order theory of 3d gravity. We show that the space of quadratic generators contains diffeomorphisms but also a notion of dual diffeomorphisms, which together form either a double Witt or centreless BMS$_3$ algebra. The relationship with the usual asymptotic symmetry algebra relies on a duality between the null and angular directions, which is possible thanks to the existence of the dual diffeomorphisms.
\end{abstract}

\maketitle

~
\vspace{-1.4cm}
\section{Motivations}
\vspace{-0.4cm}

Three-dimensional gravity has undoubtedly played an important role in the understanding of infinite-dimensional boundary symmetry algebras, and their relationship with holography, boundary dynamics, and black hole physics. Since the seminal results on asymptotically AdS$_3$ \cite{Brown:1986nw,Banados:1992wn,Coussaert:1995zp,Carlip:1994gy,Strominger:1997eq,Carlip:1999cy}, notable developments have been achieved in asymptotically-flat spacetimes and are centered around the BMS algebra \cite{Ashtekar:1996cd,Barnich:2006av,Barnich:2012rz,Barnich:2013yka,Barnich:2014kra,Barnich:2015uva,Barnich:2015sca,Barnich:2015mui,Oblak:2016eij,Carlip:2019dbu}. This latter was introduced in four-dimensional gravity early on \cite{Bondi:1962px,Sachs:1962zza,Sachs:1962wk}, and has since then found very important applications \cite{Arcioni:2003xx,Arcioni:2003td,Barnich:2010eb,Strominger:2017zoo,Ashtekar:2018lor,Fotopoulos:2019vac,Ruzziconi:2020cjt}.

In parallel, efforts have also been devoted towards the understanding of symmetries associated with boundaries at finite distance \cite{Balachandran:1991dw,balach1993MCS,Balachandran:1994up,Donnelly:2016auv,Dittrich:2017hnl,Dittrich:2017rvb,Dittrich:2018xuk,Goeller:2019apd,Goeller:2019zpz,Freidel:2016bxd,Freidel:2019ees,Freidel:2020xyx,Freidel:2020svx,Freidel:2020ayo,Wieland:2017zkf,Wieland:2017cmf,Wieland:2017ksn,Wieland:2018ymr,Wieland:2020gno}. For an arbitrary boundary in a given theory of gravity, one would like to understand the most general boundary symmetry algebra, and how it reduces to the particular cases which have been studied in the literature. In three-dimensional gravity, there is a gap in this understanding, which we here propose to fill.

In the first order formulation, the most general finite distance boundary symmetry algebra is the current algebra of the group of spacetime isometries. It can also be realized asymptotically with the most general boundary conditions of \cite{Grumiller:2016pqb,Grumiller:2017sjh}. Seen as a universal enveloping algebra, this current algebra actually contains the diffeomorphisms, which can be viewed as field-dependent gauge transformations, or quadratics in the currents.

It turns out that the space of well-defined quadratics is actually two-dimensional. This fact seems to have gone unnoticed so far. It implies the existence of dual diffeomorphisms which, together with the usual ones, form at finite distance and without boundary conditions a double Witt algebra whose flat limit is a centreless $\mathfrak{bms}_3$ algebra. When computed for arbitrary vector fields in e.g. Bondi gauge, the dual diffeomorphism charge is equal to the diffeomorphism one with the null and tangential directions exchanged. This explains why, even when considering tangential vector fields alone, these charges can reproduce asymptotic symmetry algebras at finite distance, in a manner reminiscent of symplectic symmetries \cite{Compere:2014cna,Compere:2015knw}.

In summary, our analysis explains how the double Witt and $\mathfrak{bms}_3$ algebras arise from the current algebra. For generality we perform this study on the most general first order theory \cite{Mielke:1991nn,Baekler:1992ab,Blagojevic:2004hj,Cacciatori:2005wz,Giacomini:2006dr,Blagojevic:2006jk,Blagojevic:2006hh,Blagojevic:2005pd,Blagojevic:2013bu,Cvetkovic:2018ati,Klemm:2007yu,Ning:2018gfm,Peleteiro:2020ubv}, which allows for the presence of curvature, torsion, and three independent central charges.

\vspace{-0.4cm}
\section{Lagrangian and phase space}
\vspace{-0.4cm}

Our starting point is the most general first order Lorentz-invariant theory constructed with a triad and a connection variable. This is described by the so-called Mielke--Baekler Lagrangian \cite{Mielke:1991nn}
\be\nonumber
L
&=\f{\sigma_0}{3}e\wedge[e\wedge e]+2\sigma_1e\wedge F\cr
&\pe+\sigma_2\omega\wedge\left(\de\omega+\f{1}{3}[\omega\wedge\omega]\right)+\sigma_3e\wedge\de_\omega e.
\ee
Its equations of motion can be written in the form
\be\label{EOMs}
2F+p[e\wedge e]\approx0,
\q
2\de_\omega e+q[e\wedge e]\approx0,
\ee
which reveals sources of Lorentzian curvature and torsion measured respectively by $p$ and $q$. These parameters are determined by the couplings via $\sigma_0=p\sigma_1+q\sigma_3$ and $\sigma_3=p\sigma_2+q\sigma_1$. The solution to the torsion equation of motion is $\omega=\Gamma-qe/2$, where $\Gamma(e)$ is the torsionless Levi--Civita connection. This can then be used to find the second order form of the equations of motion, which is $R_{\mu\nu}+2\lambda g_{\mu\nu}=0$ with $\lambda=(p+q^2/4)\in\mathbb{R}$.

Sticking to the first order formulation, we find that the pre-symplectic potential, which is the starting point for the covariant phase space analysis, takes the simple form $\theta=2\sigma_1\delta\omega\wedge e+\sigma_2\delta\omega\wedge\omega+\sigma_3\delta e\wedge e$.

The theory is invariant under the internal Lorentz transformations $\delta^\text{j}$ and translations $\delta^\text{t}$, which act as
\be\label{infinitesimal symmetries}
\begin{matrix*}[l]
\delta^\text{j}_\alpha e=[e,\alpha],&\q&\delta^\text{j}_\alpha\omega=\de_\omega\alpha,\cr
\delta^\text{t}_\phi e=\de_\omega\phi+q[e,\phi],&\q&\delta^\text{t}_\phi\omega=p[e,\phi].
\end{matrix*}
\ee
The commutators of these internal gauge transformations can be represented by the six-dimensional algebra
\be\label{JT algebra}
\begin{matrix}
[J^i,T^j]={\eps^{ij}}_kT^k,
\q
[J^i,J^j]={\eps^{ij}}_kJ^k,\cr
[T^i,T^j]={\eps^{ij}}_k(pJ^k+qT^k),
\end{matrix}
\ee
whose Casimir operators are
\be\label{Casimirs}
C_1=T^iT_i+pJ^iJ_i,
\q
C_2=J^iT_i+T^iJ_i-qJ^iJ_i.
\ee
On-shell of \eqref{EOMs}, diffeomorphisms $\delta^\text{d}_\xi=\pounds_\xi$ can be written as field-dependent gauge transformations in the form
\be\label{d=j+t}
\delta^\text{d}_\xi\approx\delta^\text{j}_{\xi\ip\omega}+\delta^\text{t}_{\xi\ip e}.
\ee
This follows from the topological nature of the theory.

\vspace{-0.4cm}
\section{Quasi-local algebra and dual diffeomorphisms}
\vspace{-0.4cm}

Using the covariant phase space formalism and considering field-independent gauge parameters, the charges of Lorentz transformations and translations are found to be
\be\nonumber
\J(\alpha)=2\soint\alpha(\sigma_1e+\sigma_2\omega),
\q
\T(\phi)=2\soint\phi(\sigma_1\omega+\sigma_3e),
\ee
where all the boundary integrals are on $\varphi\in S^1$. These charges form the centrally-extended current algebra
\be\nonumber
\lb\J(\alpha),\T(\phi)\rb&=\T([\alpha,\phi])-2\sigma_1\soint\alpha\de\phi,\cr
\lb\J(\alpha),\J(\beta)\rb&=\J([\alpha,\beta])-2\sigma_2\soint\alpha\de\beta,\cr
\lb\T(\phi),\T(\chi)\rb&=p\J([\phi,\chi])+q\T([\phi,\chi])-2\sigma_3\soint\phi\de\chi.
\ee
After defining the generator $\P\coloneqq\T-q\J/2$, this can be rewritten in the more familiar form
\be\nonumber
\lb\J(\alpha),\P(\phi)\rb&=\P([\alpha,\phi])-c_1\soint\alpha\de\phi,\cr
\lb\J(\alpha),\J(\beta)\rb&=\J([\alpha,\beta])-c_2\soint\alpha\de\beta,\cr
\lb\P(\phi),\P(\chi)\rb&=\lambda\left(\J([\phi,\chi])-c_2\soint\phi\de\chi\right),
\ee
where the central charges are now $c_1=2\sigma_1-q\sigma_2$ and $c_2=2\sigma_2$. We will however continue to work with $\T$ for the moment, as it leads to a simpler expression for the diffeomorphisms via \eqref{d=j+t}. The $(\J,\T)$ current algebra is well-defined at any location in the bulk. The corresponding universal enveloping algebra, built with arbitrary field-dependent smearing parameters, describes in principle all the symmetries of the theory. This includes in particular the diffeomorphisms defined as \eqref{d=j+t}.

Field-dependent gauge transformations are generically not integrable without imposing extra conditions on the gauge parameters and/or the dynamical fields. In particular, when viewing diffeomorphisms as field-dependent gauge transformations as in \eqref{d=j+t}, one finds the familiar non-integrable contribution $-\xi\ip\theta$ to their charge $\slashed{\delta}\D(\xi)$. Without placing boundary conditions on $(e,\omega)$, the diffeomorphisms can therefore be made integrable by considering tangent vector fields. Their charge can then be written as
\be\nonumber
\D(\xi)=\J(\xi\ip\omega)+\T(\xi\ip e),
\ee
which mirrors \eqref{d=j+t}. Let us momentarily focus on tangent vector fields, before relaxing this requirement later on.

It is now natural to ask if, aside from the diffeomorphisms, there is another combination of field-dependent gauge transformations which can be made integrable and consistently defined on phase space, and if so what the properties of the corresponding charge is. The answer is affirmative, and the object fulfilling these requirements is the dual diffeomorphism charge. For tangent vector fields, it is defined as
\be\nonumber
\D^*(\xi)\coloneqq p\J(\xi\ip e)+q\T(\xi\ip e)+\T(\xi\ip\omega).
\ee
The proof leading to  this charge is given for completeness in the appendix. It boils down to showing that the space of well-defined field-dependent gauge transformations is two-dimensional and can be parametrized by the diffeomorphism charge and its dual. The explicit expression for these charges, which we will need later on, is
\be\label{D and D*}
\begin{matrix*}[l]
\phantom{^*}\D(\xi)\ds=\soint~2\sigma_1(\xi\ip\omega)e+\sigma_2(\xi\ip\omega)\omega+\sigma_3(\xi\ip e)e,\cr
\D^*(\xi)\ds=\soint~2\sigma_3(\xi\ip\omega)e+\sigma_1(\xi\ip\omega)\omega+\sigma_0(\xi\ip e)e,
\end{matrix*}
\ee
where we should recall that in order to write these charges we have assumed that $\xi$ is tangent and field-independent. Notice that the dual diffeomorphisms cannot be obtained from Chern--Simons theory, since there the only field-dependent gauge transformation which can be written is $\F(\xi\ip A)$, and corresponds to the diffeomorphisms.

With the definition of $\D^*$ in terms of field-dependent gauge transformations, we can compute the action of the corresponding bulk symmetry on the fields using \eqref{infinitesimal symmetries}. On-shell of \eqref{EOMs} this gives
\be\label{action of D*}
\delta^\text{d*}_\xi e\approx\pounds_\xi\omega+q\pounds_\xi e,
\q
\delta^\text{d*}_\xi\omega\approx p\pounds_\xi e.
\ee
This action is therefore geometrical, just as the combination \eqref{d=j+t} giving the action of the diffeomorphisms $\D$. As surprising as it may seem, one can explicitly check that the transformations $\delta^\text{d*}_\xi$ are indeed symmetries of the theory. This is in fact to be expected since we have obtained $\D^*$ in terms of field-dependent gauge transformations, and \textit{all} such transformations are symmetries of the Lagrangian regardless of the nature of the gauge parameters in \eqref{infinitesimal symmetries}. From the point of view of the charges of field-dependent gauge transformations, the existence of $\D^*$ is therefore just as legitimate as that of $\D$. We can further argue in this direction using the Sugawara construction.

Seen as a universal enveloping algebra, the current algebra formed by $(\J,\T)$ actually \textit{contains} the diffeomorphisms, which can be reconstructed as quadratics in the currents. This is the essence of the Sugawara construction. This latter enables to obtain, starting from the current algebra, a Witt algebra from generators which are quadratic in the currents \footnote{The Sugawara construction can then be twisted in order to introduce a central extension and promote the Witt algebra to a Virasoro one.}. Although in three-dimensional gravity the Sugawara construction has only been used to describe the diffeomorphisms $\D$, we can now show that it also consistently leads to the dual diffeomorphisms $\D^*$.

For this, let us switch to the Fourier representation, where the currents are denoted $(\J^i_n,\T^i_n)$ with $n\in\mathbb{Z}$. The generators \eqref{JT algebra} are the zero-modes $n=0$. We then consider the quadratic generators
\be\nonumber\textstyle
\begin{matrix}
\Q_n^1\coloneqq\sum_k(\J^i_{n+k}\T^i_{-k}+\T^i_{n+k}\J^i_{-k}),\cr
\Q_n^2\coloneqq\sum_k\T^i_{n+k}\T^i_{-k},
\q
\Q_n^3\coloneqq\sum_k\J^i_{n+k}\J^i_{-k},
\end{matrix}
\ee
where the sums run over $\mathbb{Z}$. The Sugawara construction gives the Fourier expression of the diffeomorphisms and their dual in terms these quadratics in the currents. We find
\be\nonumber
\D_n&=\f{1}{4(\sigma_2\sigma_3-\sigma_1^2)}\big(\sigma_1\Q^1_n-\sigma_2\Q^2_n-\sigma_3\Q^3_n\big),\cr
\D^*_n&=\f{1}{4(\sigma_2\sigma_3-\sigma_1^2)}\big(p\sigma_1\Q^3_n-p\sigma_2\Q^1_n+(\sigma_1-q\sigma_2)\Q^2_n\big),
\ee
which as expected can be shown to match the previous definitions of these generators with $\xi_\para=e^{-\i n\varphi}\partial_\varphi$. We see that the dual diffeomorphisms $\D^*_n$ are therefore on the same footing as $\D_n$. Together, the diffeomorphisms and their dual exhaust the possibilities for constructing quadratic generators forming a well-defined algebra among themselves and with the currents.

The existence of the dual diffeomorphisms can also be traced back to the fact that the Sugawara construction relies on the quadratic Casimir operators of the global algebra underlying the current algebra of boundary symmetries \cite{DiFrancesco:639405}. In three-dimensional gravity this algebra always admits two Casimirs, which in the most general formulation studied here are given by \eqref{Casimirs}. Comparing this with the above Sugawara expressions for the quadratic generators reveals that they are given by
\be\nonumber
\D\propto\sigma_2\widetilde{C}_1-\sigma_1\widetilde{C}_2,
\q
\D^*\propto(\sigma_1-q\sigma_2)\widetilde{C}_1-p\sigma_2\widetilde{C}_2,
\ee
where the tilde denotes the lift of the Casimirs to the universal enveloping algebra. The two-dimensionality of the space of well-defined quadratic operators matches the fact that the underlying algebra has two Casimirs. This is a non-trivial consistency check, as one can show that it is only the quadratics forming a stable algebra with the currents which can be written in terms of the Casimirs.


Sticking to tangent vectors, we can now compute the algebra of the diffeomorphisms and their dual, to find
\be\nonumber
\lb\D^*(\xi),\D(\zeta)\rb&=-\D^*([\xi,\zeta]),\cr
\lb\D(\xi),\D(\zeta)\rb&=-\D([\xi,\zeta]),\cr
\lb\D^*(\xi),\D^*(\zeta)\rb&=-p\D([\xi,\zeta])-q\D^*([\xi,\zeta]).
\ee
Remarkably, this reflects exactly the $(\J,\T)$ current algebra from which we started, and can therefore be thought of as its representation in terms of vector fields instead of Lie algebra elements. However, since the vector fields are tangent we obtain no central extensions. This algebra can now be put in a more suggestive form by redefining the dual generator as $\A\coloneqq\D^*-q\D/2$, leading to
\be\label{AD algebra}
\lb\A(\xi),\D(\zeta)\rb&=-\A([\xi,\zeta]),\cr
\lb\D(\xi),\D(\zeta)\rb&=-\D([\xi,\zeta]),\\
\lb\A(\xi),\A(\zeta)\rb&=-\lambda\D([\xi,\zeta]),\nonumber
\ee
which reflects the $(\J,\P)$ current algebra. In the case $\lambda=0$, which can be achieved for $q^2=-4p$, this is the centreless $\mathfrak{bms}_3$ algebra. When $\lambda\neq0$, the redefinition $\D^\pm\coloneqq(\D\pm\lambda^{-1/2}\A)/2$ reveals as usual the direct sum of two Witt (or centreless Virasoro) algebras. Note that this also works in the dS case $\lambda<0$.

This is the main result of the present letter, namely the construction, in the most general first order theory of three-dimensional gravity, of the quadratic generators in the universal enveloping algebra of boundary symmetries. These are the diffeomorphisms and their dual, which at any finite distance form a double Witt algebra. We note that the only related result where a $\mathfrak{bms}_3$ algebra was built from the current algebra using two independent quadratics seems to be \cite{Caroca:2017onr}, which however did not provide a gravitational interpretation of this construction.

\vspace{-0.4cm}
\section{Relation with asymptotic charges and algebra}
\vspace{-0.4cm}

It is now enlightening to compare our construction with the derivation of the asymptotic double Virasoro algebra and its flat limit. For this, we consider the Bondi gauge line element \cite{Barnich:2012aw}
\be\nonumber
\de s^2=(\mathscr{M}-\lambda r^2)\de u^2-2\de u\,\de r+\mathscr{N}\,\de u\,\de\varphi+r^2\de\varphi^2,
\ee
where the two free functions $(\mathscr{M},\mathscr{N})$ are independent of $r$ and satisfy $\partial_u\mathscr{M}=\lambda\mathscr{N}'$ and $\partial_u\mathscr{N}=\mathscr{M}'$, where prime is the angular derivative. The on-shell asymptotic Killing vectors are $\xi=(\xi^u,\xi^r,\xi^\varphi)$ with
\be\nonumber
\xi^u=f,
\q
\xi^r=f''-rg'-\f{\mathscr{N}f'}{2r},
\q
\xi^\varphi=g-\f{f'}{r},
\ee
where $(f,g)$ are independent of $r$ and satisfy $\partial_ug=\lambda f'$ and $\partial_uf=g'$. These diffeomorphisms preserve the family of metrics, and change their parameters as
\be\nonumber
\delta^\text{d}_\xi\mathscr{M}&=g\mathscr{M}'+2\mathscr{M}g'-2g'''+\lambda(2\mathscr{N}f'+f\mathscr{N}'),\cr
\delta^\text{d}_\xi\mathscr{N}&=f\mathscr{M}'+2\mathscr{M}f'-2f'''+2\mathscr{N}g'+g\mathscr{N}'.
\ee
Since these vector fields are field-dependent and also non-tangent, their diffeomorphism charge should be computed from the variational expression $\slashed{\delta}\D(\xi)$, and in particular contains the piece $-\xi\ip\theta$. Using a triad $e$ for the above metric, and the connection $\omega=\Gamma-qe/2$ solving the torsion equation \footnote{For completeness we give the triad and connection components in the appendix.}, the diffeomorphism charge is then found to be integrable and given by
\be\nonumber
\D_\text{B}(\xi)=\f{1}{2}\soint f(c_1\mathscr{M}+\lambda c_2\mathscr{N})+g(c_1\mathscr{N}+c_2\mathscr{M})+\O(r^{-1}),
\ee
where the subscript stands for Bondi. The subleading term is exact and given by $-f'(c_1\mathscr{N}+c_2\mathscr{M})/r$. At null infinity, which is reached for $r\to\infty$, we can separate the charge into its null and angular components by writing $\D_\text{B}^\infty(\xi)=\E(f)+\L(g)$, and we find the algebra \cite{Adami:2020xkm,Geiller:2020edh}
\be\nonumber
\lb\E(f),\L(g)\rb&=-\E([f,g])+c_1\soint fg''',\cr
\lb\L(g_1),\L(g_2)\rb&=-\L([g_1,g_2])+c_2\soint g_1g_2''',\cr
\lb\E(f_1),\E(f_2)\rb&=-\lambda\left(\L([f_1,f_2])-c_2\soint f_1f_2'''\right).
\ee
Upon redefining the generators this can be put in the form of a double Virasoro algebra with central charges $c^\pm=6(\lambda^{-1/2}c_1\pm c_2)$. In the flat limit the above brackets reproduce that of the centrally-extended algebra $\mathfrak{bms}_3$ with two independent central charges.

We can observe that this algebra of asymptotic charges reflects the $(\J,\P)$ current algebra, and in particular has the same central extensions. As an algebra of vector fields, and up to the central extensions, it is evidently the same algebra as \eqref{AD algebra}, although this latter was obtained from a very different construction, i.e. using tangent vectors and the dual diffeomorphism charge. This suggests however that the angular and null pieces of the asymptotic Bondi charge can be reproduced respectively by the tangent diffeomorphism charge and its dual. Evaluating these two charges on the vector $\xi_\para=(0,0,h)$ using \eqref{D and D*} and the triad and connection components given in the appendix, we discover that this is indeed the case, as
\be\nonumber
\D(\xi_\para)\equiv\L(h)&=\f{1}{2}\soint h(c_1\mathscr{N}+c_2\mathscr{M}),\cr
\A(\xi_\para)\equiv\E(h)&=\f{1}{2}\soint h(c_1\mathscr{M}+\lambda c_2\mathscr{N}).
\ee
Although the first identification is perhaps not surprising, since $\L$ is after all the tangential part of the Bondi diffeomorphism charge, the second one is more unexpected. It shows that the non-tangential $u$ component of the Bondi gauge diffeomorphism charge $\D_\text{B}^\infty$ can be written as the dual diffeomorphism charge evaluated for a tangential vector field. In fact, one can show that an even stronger result holds. For an arbitrary vector field, evaluating the charges $\slashed{\delta}\A(\xi)$ and $\slashed{\delta}\D(\xi)$ in Bondi gauge shows that $\A(\xi^u,\xi^r,\xi^\varphi)=\D(\xi^\varphi,\xi^r,\lambda\xi^u)$. In terms of the symmetry transformations of the fields, this means in particular that the action of $\A$ in the tangent angular direction $\varphi$ is equivalent to a diffeomorphisms in the null direction $u$, i.e. that $\delta^\text{a}_{(0,0,h)}\approx\delta^\text{d}_{(h,0,0)}$. This can indeed be shown to hold by using \eqref{action of D*} and the triad and the on-shell connection components given in the appendix.

Heuristically, regarding the information captured by $\A$ everything happens as if the Cauchy slice $\Sigma$ defining the symplectic structure was redefined so as to render the $u$ direction tangential. What is also remarkable is that the reconstruction of $\D_\text{B}^\infty$ using $\D$ and $\A$ does not require to discuss boundary conditions, which in principle are required to deal with the $-\xi\ip\theta$ piece. The information along $u$ contained in this non-integrable piece has been repackage in the charge $\A$. Clearly, this mechanism is here possible because the radial component $\xi^r$ does not contribute to the asymptotic charge. Going back to the previous heuristic picture, since the $r$ direction can never be made tangential by a change of slicing, it is reasonable to expect that its contribution cannot be captured by $\D$ nor $\A$ evaluated on tangential vectors. The fact that our construction of \eqref{AD algebra} does not accommodate the radial direction is also the reason for which this algebra has no central extensions. It is however possible to perform a so-called twist of the Sugawara construction, consisting in a shift of the quadratic generators by a linear term, and thereby introduce three independent central charges in the algebra \eqref{AD algebra} \cite{Geiller:2020edh}. We keep the study of these central extensions and of the radial direction for future work.

The fact that we have recovered the asymptotic symmetry algebra at any arbitrary location in the bulk is reminiscent of the so-called symplectic nature of these symmetries \cite{Compere:2014cna,Compere:2015knw}. This property follows form the absence of bulk degrees of freedom in the theory, and relies on the fact that the diffeomorphism charge is actually independent of the radial coordinate $r$. Although this is true in the metric formulation, here, in the first order formulation, it is not manifest because of the presence of the subleading $\O(r^{-1})$ piece in the charge $\D_\text{B}$. This contribution can however be removed by replacing the Lie derivative with the Kosmann derivative \cite{Jacobson:2015uqa}. This latter is defined as $\mathscr{K}_\xi\coloneqq\pounds_\xi+\delta^\text{j}_k$ with a field-dependent parameter given by $2k^i\coloneqq-{\eps^i}_{jk}g^{\mu\nu}e^j_\mu\pounds_\xi e^k_\nu$, and is such that $\mathscr{K}_\xi e=0$ when $\xi$ is Killing. The charge of the Lorentz transformation is integrable and cancels exactly the subleading term in $\D_\text{B}$, so that the charge associated with the Kosmann derivative is therefore $r$-independent.

\vspace{-0.4cm}
\section{Perspectives}
\vspace{-0.4cm}

In this letter we have studied the finite distance symmetry algebra of the most general first order theory of three-dimensional gravity, and explained how aspects of asymptotic symmetry algebras are embedded in it. This relies on the construction of the second order charges in the universal enveloping algebra of the $(\J,\T)$ current algebra. We have shown that the existence of two (and only two) well-defined quadratic charges, namely the diffeomorphism $\D$ and its dual $\D^*$, can be understood in terms of the Sugawara construction and its dependency on the Casimirs of the global part of the current algebra. Focusing on tangent vector fields, which enables to work without imposing boundary conditions, we have shown that these charges form a double Witt algebra which in the flat limit reduces to a centreless $\mathfrak{bms}_3$ algebra. The absence of central extensions comes from the tangentiality of the vector fields, and can be lifted by considering a twisted Sugawara construction. This is outlined in \cite{Geiller:2020edh} and will be the focus of future work.

Our construction has raised a natural question: Is it possible to understand the asymptotic double Virasoro and $\mathfrak{bms}_3$ algebras in terms of the diffeomorphisms and their dual? We have shown that the answer is positive but subtle. The angular part of the asymptotic Bondi gauge charge is captured by a tangent diffeomorphism at any finite distance in the bulk. More surprisingly, the null part of the asymptotic Bondi gauge charge is captured by the dual diffeomorphism evaluated on a tangent vector field. This means that the null direction of a usual diffeomorphism can be understood as a tangential direction from the viewpoint of the dual diffeomorphism. This is reminiscent of \cite{Adami:2020ugu,ruzziconi2020conservation}, where a change of vector field basis is used to render any diffeomorphism integrable.

This opens up interesting directions. At finite distance, it is now necessary to study the diffeomorphism and its dual for arbitrary non-tangent vector fields. This requires the discussion of integrability and boundary conditions, whose relation with the twisted Sugawara construction must also be spelled out. This is a necessary generalization in order to obtain central extensions. Considering arbitrary vector fields will potentially extend the finite distance symmetry algebra, and a question is therefore whether this can also be realized asymptotically. Another crucial question is if and how these dual charges are related to those studied in four-dimensional gravity in \cite{Godazgar:2018dvh,Godazgar:2018qpq,Godazgar:2019dkh,Kol:2019nkc,Godazgar:2020gqd,Godazgar:2020kqd,Oliveri:2020xls}. Other questions concern the role of the dual diffeomorphisms in the quantization of the theory and the construction of representations of its symmetry algebra.

\vspace{-0.4cm}
\section{Appendix 1: Well-definiteness of $\boldsymbol{\D^*}$}
\vspace{-0.4cm}

To introduce the dual diffeomorphisms, we start with the most general linear combination of field-dependent gauge transformations obtained by contracting a vector field with a connection or a triad field. This is
\be\nonumber
\slashed{\delta}\G(\xi)=a\slashed{\delta}\J(\xi\ip e)+b\slashed{\delta}\J(\xi\ip\omega)+c\slashed{\delta}\T(\xi\ip e)+d\slashed{\delta}\T(\xi\ip\omega).
\ee
 Writing out explicitly these field-dependent charges, it is not hard to see that they can be made integrable without restricting the dynamical fields if we impose the algebraic relation $b\sigma_1+d\sigma_3-a\sigma_2-c\sigma_1=0$ and consider tangent vector fields. With integrability being achieved, we can then compute the brackets of these new charges with the diffeomorphisms, the Lorentz transformations, and the translations. We then find that all these brackets are well-defined, aside from
\be\nonumber
\lb\G(\xi),\T(\phi)\rb=
&-a\J(\pounds_\xi\phi)-c\T(\pounds_\xi\phi)\cr
&-2(b\sigma_3+d\sigma_0-a\sigma_1-c\sigma_3)\soint[\xi\ip e,\phi]\omega,
\ee
which does not close because of the field-dependent piece. This latter is problematic because it is not the charge of an integrable gauge transformation. Moreover, when considering iterated Poisson brackets it leads to higher and higher order terms in the fields. Cancelling this piece by imposing the vanishing of its coefficient, and choosing for convenience the normalization $d=1$, we find $a=p$ and $c=b+q$. We therefore obtain the well-defined charges $\G(\xi)=p\J(\xi\ip e)+q\T(\xi\ip e)+\T(\xi\ip\omega)+b\D(\xi)$. Finally, we can take $b=0$ and consider $\D^*$ and $\D$ as the independent charges. In summary, starting from the four-dimensional space of field-dependent gauge transformations $\slashed{\delta}\G(\xi)$, we have imposed two conditions in order to achieve integrability and obtain a stable algebra. We are therefore left with a two-dimensional space, which can be parametrized by the independent generators $\D^*$ and $\D$.

\vspace{-0.4cm}
\section{Appendix 2: Triad and connection components}
\vspace{-0.4cm}

In order to make this letter self-contained, we give a choice of triad compatible with the line element in Bondi gauge. The internal metric and the triad are
\be\nonumber
\eta_{ij}=
\begin{pmatrix}
0&1&0\\
1&0&0\\
0&0&1
\end{pmatrix},
\q
\everymath={\displaystyle}
e^i_\mu=
\begin{pmatrix}
(\mathscr{M}-\lambda r^2)/2&1&0\\
-1&0&0\\
\mathscr{N}/2&0&r
\end{pmatrix}.
\ee
The connection solving the torsion equation is
\be\nonumber
\everymath={\displaystyle}
\omega^i_\mu=
\begin{pmatrix}
\lambda\mathscr{N}/2&0&\lambda r\\
0&0&0\\
(\mathscr{M}-\lambda r^2)/2&1&0
\end{pmatrix}
-\f{q}{2}e^i_\mu.
\ee
The lines and columns correspond to the $\mu$ and $i$ indices.

\bibliography{Biblio.bib}

\end{document}